\documentclass[12pt]{iopart}

\usepackage{amssymb}
\usepackage{amsfonts}
\usepackage{graphicx}

\begin{document}

%\begin{frontmatter}
%\date{\today}
%\title[Short title]{Symplectic integration of deviation vectors and chaos
%  determination. Application to the restricted three-body problem.}
%\author{A.-S. Libert, C. Hubaux, T. Carletti}

\title[Symplectic integration of deviation vectors and chaos
  determination.]{Symplectic integration of deviation vectors and chaos
  determination. Application to the H\'enon-Heiles model and to the
  restricted three-body problem.}%\tnoteref{label1}} 
%\tnotetext[label1]{}
\author{A.-S. Libert, Ch. Hubaux and T. Carletti}% \corref{cor1}}%\fnref{label2}}
%\ead{anne-sophie.libert@fundp.ac.be}
%\author{Charles Hubaux}
%\ead{charles.hubaux@fundp.ac.be}
%\author{Timoteo Carletti}
%\ead{timoteo.carletti@fundp.ac.be}
%\fntext[label2]{}
%\cortext[cor1]{}
\address{Department of Mathematics FUNDP, 8 Rempart de la Vierge, B-5000 Namur, Belgium}%\fnref{label3}}

\eads{\mailto{anne-sophie.libert@fundp.ac.be}, \mailto{charles.hubaux@fundp.ac.be}, \mailto{timoteo.carletti@fundp.ac.be}}
%\fntext[label3]{}

\begin{abstract}
In this work we propose a new numerical approach to distinguish between
  regular and chaotic orbits in Hamiltonian systems, based on the simultaneous
  integration of both the orbit and the deviation vectors using a symplectic
  scheme, hereby called {\em global symplectic integrator}.
In particular, the proposed method allows us to recover the correct orbits
character with very large integration time steps, small energy losses and
short CPU times. To illustrate the numerical performances of the 
  global symplectic integrator we will apply it to two well-known and widely
  studied problems: the H\'enon-Heiles model and the restricted three-body
  problem. 
\end{abstract}

%\begin{keyword}
%Stability; Chaos indicator; Symplectic integrator; Hamiltonian system  
%\end{keyword}

%\end{frontmatter}

\maketitle
\section{Introduction}
\label{sec:intro}

Hamiltonian systems exhibit phase spaces where regular and chaotic orbits do
coexist. This fact makes the problem of the numerical characterization of
regular and chaotic motion an hard task, notably in systems with many degrees of freedom. For this reason scientists developed fast and accurate tools to obtain
information about the chaotic versus regular nature of the orbits of such systems, and efficiently characterize large domains in their phase space as ordered or (weakly) chaotic.

These methods can be roughly divided into two major groups: {\em
  Lyapunov--like} methods, i.e. methods based on the study of the evolution of deviation vectors for a given orbit; the methods
FLI~\cite{Froe1997}, \texttt{MEGNO}~\cite{cincotta2000},
\texttt{SALI}~\cite{S01} and \texttt{GALI}~\cite{SBA07} belong to
this first class. And {\em Fourier--like} methods based on the determination
of the frequencies of the spectrum of some observable related to a given
orbit, for instance \texttt{FMA}~\cite{Laskar1992,Laskar1993} or the
Spectral method~\cite{GB2001}. In the present work we will be interested
in the former class, in particular to the \texttt{SALI} chaos detection
technique. 

To numerically compute such an indicator, one needs to have a good orbit
determination, but also to integrate the evolution of two deviation
vectors. It is well-known that symplectic integration schemes outperform the
non--symplectic ones, when compared using the same order of accuracy and the same
integration step size. This is mainly due to the very good energy conservation
properties, but also to other first integrals. As a consequence, larger step sizes are allowed while still keeping a reasonably small energy loss. Thus they allow us to enlarge
considerably the time span of the numerical simulation, without degradating
the goodness of the numerical results, that is why symplectic integrators turn
out to be 
essential for the long-term evolution of a dynamical system. 

In the same way,
chaos detection techniques could also benefit from the use of symplectic
integrators. The aim of this paper is to show that the symplectic integration
of the 
deviation vectors can improve the ability of the previous chaos indicators in
the characterization of regular and chaotic orbits by correctly identifying the orbit behavior, using a larger integration step size. Because
our method propose to integrate simultaneously both the orbit and the
deviation vectors, using a symplectic scheme, we hereby name this method {\em
  global symplectic integrator}.

More precisely, we will accurately determine, using a
symplectic scheme and a small enough integration
time step $\tau$, the character, i.e. the regular or chaotic behavior,
of a large number, $N_{tot}$, of orbits using one of the aforementioned chaos
detectors. Then, all the 
previous orbits will be reanalyzed using both symplectic and non-symplectic integrators schemes, but
with 
larger and larger time steps. We will show that the use of the global
symplectic  
integrator will allow us to recover a large percentage
of orbits' characters with very large time steps compared to
the non--symplectic one and using small amount of CPU time.

Our symplectic method will be applied to the classical H\'enon--Heiles model, and also to a well-known problem of celestial mechanics: the restricted three-body problem, where the possibility of using large integration step sizes is essential for the study of secular resonances.

The paper is organized as follows. In Section~\ref{sec:themethod}, we describe
our method for integrating symplectically the deviation vectors used in
the chaos detection techniques. For the sake of completeness, we
present, in Section~\ref{sec:sali}, a brief introduction to the \texttt{SALI}
chaos indicator. The results of our method applied to the
H\'enon--Heiles system are presented in Section~\ref{ssec:hh}, while Section~\ref{ssec:3b} will be devoted to
  present an application to the restricted
three-body problem. Finally in Section~\ref{sec:ccl} we will summarize our
findings and we will conclude.

\section{The method}
\label{sec:themethod}

Let us consider a generic Hamiltonian vector field
\begin{equation}
  \label{eq:hamsyst}
  \dot{\vec{x}}=J\nabla_x H(\vec{x})\, ,
\end{equation}
where $\vec{x}=(\vec{p},\vec{q})\in{R}^{2n}$ is the momentum--position vector in the
phase 
space, $H(\vec{x})$ is the Hamilton function describing the system and $J=\left(
  \begin{array}{cc}
    0_n & -1_n\\1_n & 0_n
  \end{array}
\right)$ is
the standard constant symplectic matrix. The solution of~(\ref{eq:hamsyst})
with initial datum $\vec{x}_0$ can be formally written as
\begin{equation}
  \label{eq:fromsol}
  \varphi(t)=e^{tL_H}\vec{x}_0\, ,
\end{equation}
where $L_H$ is the Lie operator, i.e. for all smooth function defined in the
phase space we have $L_Hf=\{H,f\}$, being $\{\cdot,\cdot\}$ the Poisson
  bracket.

Given a deviation vector $\vec{v}$, its time evolution is described by the
{\em tangent map}:
\begin{equation}
  \label{eq:tangmap}
  \frac{d\vec{v}}{dt}=M\left(\varphi(t)\right)\vec{v}\, , 
\end{equation}
where $M$ is the Jacobian of the Hamiltonian vector field~(\ref{eq:hamsyst}),
namely $M=J\nabla_x^2H$, evaluated on the solution $\varphi(t)$.

Because
$\nabla_x^2H$ is a symmetric matrix, the matrix $M$ 
is an Hamiltonian one, i.e. $M^TJ+JM=0$. Hence also the vector
field~(\ref{eq:tangmap}) is Hamiltonian, with Hamiltonian function
\begin{equation}
  \label{eq:Kham}
 K(\vec{v})=\frac{1}{2}\vec{v}^T S\vec{v}\, ,
\end{equation}
 where $S=\nabla_x^2H$.

Let us now assume that the function $H$ can be decomposed into the sum of two
parts each one separately integrable, for instance each one depending only on
one group of variables 
\begin{equation}
  \label{eq:HAplusB}
  H(\vec{x})=A(\vec{p})+B(\vec{q})\, .
\end{equation}
Let us observe that in the case of the restricted three-body problem
  (see Section~\ref{ssec:3b}),
  we will prefer to use a different splitting of the Hamiltonian
  function: actually the function $A$ will correspond to two
  non-interacting two-body problems that in rectangular coordinates will
  thus result a function of positions and momenta; while the remainder will be
  by 
  definition the $B$ function.

Under the above assumptions, a family of suitable
symmetric symplectic schemes has been 
presented in~\cite{LR2001}, hereby called $SABA_n$. Such methods develop the
solution given by~(\ref{eq:fromsol}) using the Baker--Campbell--Hausdorff 
formula~\cite{Bourb1972}, allowing to approximate the true solution by a
finite number of compositions of integrable symplectic maps, being the error
of the approximation a
function of the integration step size $\tau$.

More precisely one can find coefficients $(c_i)_{i=1,\dots,n+1}$ and
$(d_i)_{i=1,\dots,n}$ such that the map
\begin{equation}
  \label{eq:saban}
  SABA_{n}(\vec{x})=e^{c_1\tau L_A}e^{d_1\tau L_B}\dots e^{d_n\tau
    L_B}e^{c_{n+1}\tau L_A}e^{d_n\tau L_B}\dots e^{d_1\tau L_B}e^{c_1\tau
    L_A}(\vec{x})\, ,
\end{equation}
is symmetric under the transformation $t\mapsto -t$, is symplectic and is of
order $\mathcal{O}(\tau^{2n})$, namely there exists an Hamiltonian function
$\tilde{H}$ whose exact flow is given by~(\ref{eq:saban}) and moreover
$\tilde{H}=H+\mathcal{O}(\tau^{2n})$. 

Rewriting the deviation vector as $\vec{v}=(\vec{v}_p,\vec{v}_q)$,
i.e. identifying its components with the natural splitting of the phase space,
we can explicitely write the Hamiltonian $K$ 
as follows:
\begin{equation}
  \label{eq:K}
  K(\vec{v})=\frac{1}{2} \vec{v}_p^T
  \mathcal{A}(\vec{p})\vec{v}_p+\frac{1}{2} \vec{v}_q^T
  \mathcal{B}(\vec{q})\vec{v}_q\, ,
\end{equation}
where
\begin{equation}
  \label{eq:AetB}
  \mathcal{A}(\vec{p})=\nabla_p^2A(\vec{p})\quad and\quad
  \mathcal{B}(\vec{q})=\nabla_q^2B(\vec{q})\, .
\end{equation}
Hence the symmetric symplectic methods $SABA_n$ can be used also to integrate
the evolution of the deviation vectors.

The aim of this paper is to show that the use of a symplectic integrator to
solve both the orbit evolution and the tangent equation, i.e. to get the time evolution of the deviation
vector, can improve the capability of some widely used chaos indicators (such as \texttt{SALI}) to determine the character of the orbits in an Hamiltonian system.

Our claim will be numerically proved for the H\'enon--Heiles system, using
$SABA_4$ and $SABA_2$ symplectic schemes in comparison with the
non--symplectic $4^{th}$ order Runge--Kutta method (see
Section~\ref{ssec:hh}); while $SABA_{10}$ and Bulirsch-Stoer will be used for
the study of the
  restricted three-body problem (see Section~\ref{ssec:3b}). In the
rest of the paper the easily implementable \texttt{SALI} chaos
  detection technique will be used to numerically determine the orbit
  character. For the
sake of completeness, the next section will be devoted to a brief
introduction of the above mentioned chaos indicator.

\section{The SALI chaos indicator}
\label{sec:sali}
The Smaller ALignment Index, \texttt{SALI}~\cite{S01}, has been proved to be
an efficient 
and simple method to determine the regular or chaotic nature of orbits
in conservative dynamical systems. Thanks to its properties it has been already
successfully applied to distinguish between regular and chaotic 
motion both in symplectic maps and Hamiltonian flows
(e.g. \cite{SABV03,SESS04,MA05,BCSV2010}).

For the sake of completeness let us briefly recall the definition of the
\texttt{SALI} 
and its behavior for 
regular and chaotic orbits, restricting our attention to
 symplectic flows. The interested reader can
consult~\cite{S01,SABV04} to have a more detailed description of the method.  
To compute the 
\texttt{SALI} of a given orbit, one has to follow the time
evolution of the orbit itself and also of two linearly independent unitary
deviation vectors 
$\hat{v}_{1}(0),\hat{v}_{2}(0)$. The evolution of an orbit is given
by~(\ref{eq:fromsol}), while the evolution of each deviation
vector is given by the tangent map~(\ref{eq:tangmap}).

Then, according to~\cite{S01} the \texttt{SALI} for the given orbit is defined
by 
\begin{equation}\label{eq:SALI:2}
\texttt{SALI}(t)=\min
\left\{\left\|\hat{v}_{1}(t)+\hat{v}_{2}(t)\right\|,\left\|\hat{v}_{1}(t)-\hat{v}_{2}(t)\right\|\right\}\,
  , 
\end{equation}
where $\| \cdot \|$ denotes the usual Euclidean norm and
$\hat{v}_{i} (t)=\frac{\vec{v}_{i}(t)}{\| \vec{v}_{i}(t)\|}$, $i=1,2$, are normalized vectors.

In the case of chaotic orbits, the deviation vectors $\hat{v}_1$,
$\hat{v}_2$ eventually become aligned in the direction defined by the
maximal Lyapunov characteristic exponent~\cite{benettin80a} (LCE), and
$\texttt{SALI}(t)$ falls 
exponentially to zero. An analytical study of \texttt{SALI}'s behavior for
chaotic orbits was carried out in~\cite{SABV04} where it was shown
that
\begin{equation}
\texttt{SALI}(t) \propto e^{-(\sigma_1-\sigma_2)t}, 
\label{eq:exp}
\end{equation}
with $\sigma_1$, $\sigma_2$ being the two largest LCEs.

In the case of regular motion, on the other hand, the orbit lays on a
torus and the vectors $\hat{v}_1$, $\hat{v}_2$ eventually fall on its
tangent space, following a $t^{-1}$ time evolution, having in general
different directions. In this case, the \texttt{SALI} oscillates about non zero values
(for more details see~\cite{SABV03}). This behavior is due to the fact that
for regular orbits the norm of a deviation vector increases linearly in
time. Thus, the normalization procedure brings about a decrease of the
magnitude of the coordinates perpendicular to the torus at a rate
proportional to $t^{-1}$ and so $\hat{v}_1$, $\hat{v}_2$ eventually
fall on the tangent space of the torus.

The simplicity of \texttt{SALI}'s definition, its completely different behavior
for regular and chaotic orbits and its rapid convergence to zero in
the case of chaotic motion are the main advantages that make \texttt{SALI} an
ideal chaos detection tool. Recently a generalization of the \texttt{SALI}, the
so-called Generalized Alignment Index, \texttt{GALI}, has been
introduced~\cite{SBA07,SBA08}, which uses information of more than two
deviation 
vectors from the reference orbit. Since the advantages of \texttt{GALI} over
\texttt{SALI} become relevant in the case of multi-dimensional systems, for
the aim of the
present paper we will restrict our attention to the \texttt{SALI}.

The scheme we used to numerically compute the \texttt{SALI} consists in
integrating both the orbit and the deviation vector using 
the symmetric symplectic $SABA$ method and then compute the indicator
according to the definition~(\ref{eq:SALI:2}).

\section{Results}
\label{sec:appl}
%We are now able to present our results.
In this section, we will present the results of the application of the global
symplectic integrator to study the orbit behavior of two well-known
problems: the H\'enon--Heiles system and the restricted three-body
problem. The analysis of the former widely studied dynamical system, will
point out that chaos indicator can benefit from our method, regarding the
characterization of 
regular and chaotic motion (see Section \ref{ssec:hh}). The second application
of our method consists in a classical problem of celestial mechanics, namely
the restricted three-body problem, that will be presented in Section \ref{ssec:3b}. 

\subsection{H\'enon--Heiles system}
\label{ssec:hh}

\begin{figure}%[htbp]
\begin{center}
\includegraphics[height=10cm]{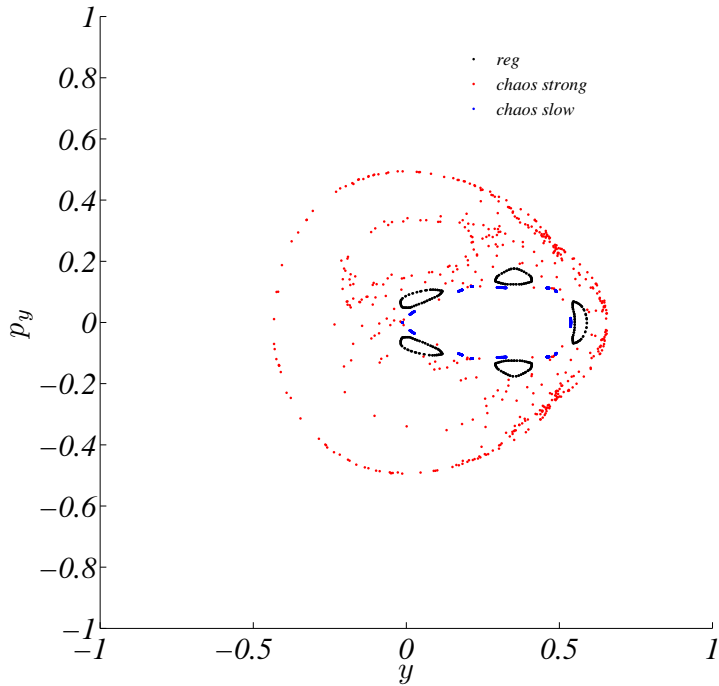}
\end{center}
\caption{H\'enon--Heiles phase space (section $x=0$). The energy has been
  fixed to $E=1/8$ 
  and three characteristic orbits have been computed on this energy level:
  a regular orbit (black) with initial data $x(0)=0$, $y(0)=0.55$, $p_x\sim
  0.2417$ and $p_y=0$, a chaotic orbit strongly diffusing (red) with initial data $x(0)=0$, $y(0)=-0.016$, $p_x\sim 0.49974$ and $p_y=0$, and a chaotic orbit slowly diffusing (blue) with initial data $x(0)=0$, $y(0)=-0.01344$, $p_x\sim 0.49982$ and $p_y=0$.} 
\label{fig:hhphasespace}
\end{figure}

%\begin{remark}[About the choice of the numerical integrators]
%To compare our results we need to use reliable symplectic and non--symplectic
%integrators with the same precision. Moreover the smaller the used times step
%size or the higher the integrator order, the better will be the energy
%preservation also for the non--symplectic scheme in a finite, a priori chosen
%integration interval. Because we would emphasize the role the symplectic
%scheme, without changing the integration interval, we decided that a $4^{th}$
%\end{remark}

The H\'enon--Heiles Hamiltonian system~\cite{HH1964} is a well-known and
studied model described by the following Hamilton function
\begin{equation}
  \label{eq:hamhh}
H(p_x,p_y,x,y)= \frac{1}{2}\left(p_x^2+p_y^2+x^2+y^2\right)+x^2y-\frac{1}{3}y^3\, .
\end{equation}
In the phase space, regular and chaotic orbits coexist (see
Figure~\ref{fig:hhphasespace}), whose character can be accurately and
rapidly determined using \texttt{SALI} and integrating both the orbit and the
deviation vectors using the $SABA_4$ scheme with a time step size
$\tau=10^{-4}$. The choice of a $4^{th}$ order integrator is motivated by a balance between reliability and computation time. Indeed, the smaller the used time step or the higher the integrator order, the better the energy preservation, but the longer the computation time. In this respect, we decided to adopt a $SABA_4$ integrator with a sufficiently small time step of $\tau=10^{-4}$.

\begin{figure}%[htbp]
\begin{center}
\includegraphics[height=7cm]{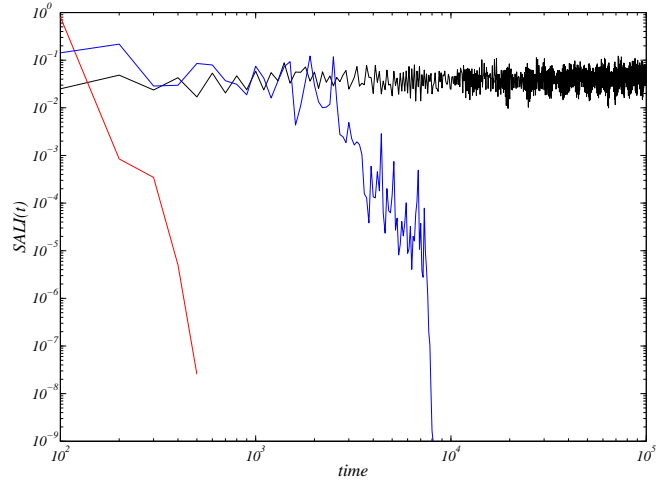}
\end{center}
\caption{Characterization of the three orbits of Figure \ref{fig:hhphasespace}: a regular orbit (black), a chaotic orbit strongly diffusing (red) and a chaotic orbit slowly diffusing (blue). The integration step size has been fixed to $\tau=10^{-4}$ and both the orbit and the deviation vectors have been numerically integrated using the $SABA_4$ method.}
\label{fig:sali3orbhh}
\end{figure}

\begin{figure}%[htbp]
\begin{center}
\includegraphics[height=7cm]{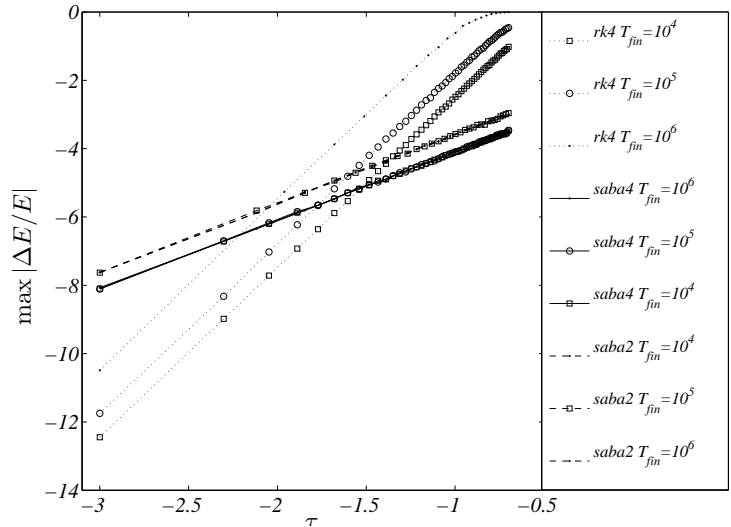}
\end{center}
\caption{Relative energy loss, $\max_{0\leq t\leq T_{fin}}|\Delta E(t) / E(t)|$,
  for $SABA_4$, $SABA_2$ and $RK4$ integrators as a function of the time
  steps. Both quantities are given in logarithmic scale. Several integration
  times are selected, $T_{fin}=\{ 10^4, 10^5,10^6\}$, as reported in the
  legend.}
\label{fig:energy}
\end{figure}

\begin{figure}[htbp]
\begin{center}
\hspace{0.5cm}\includegraphics[width=5cm]{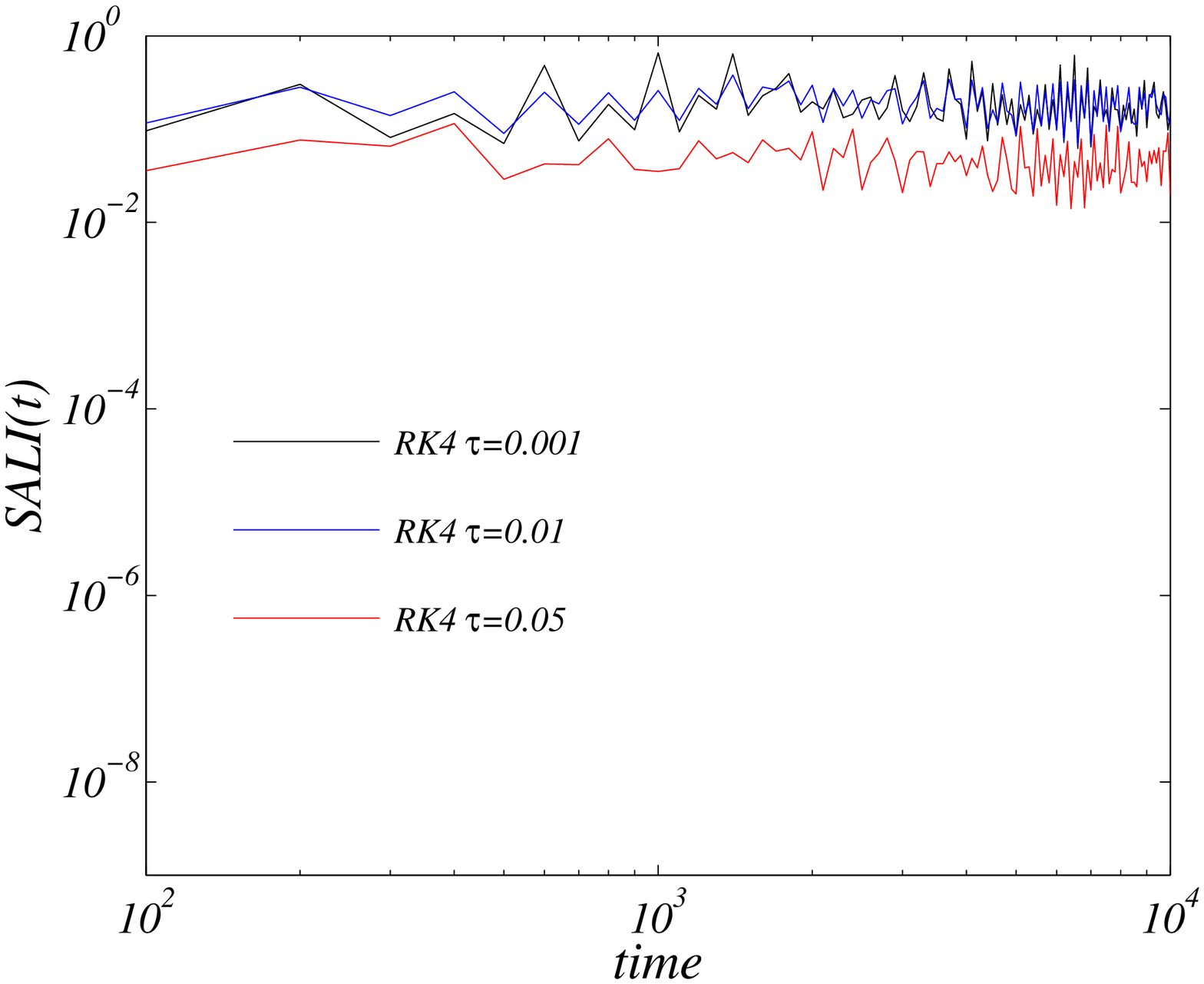}\quad \includegraphics[width=5.6cm]{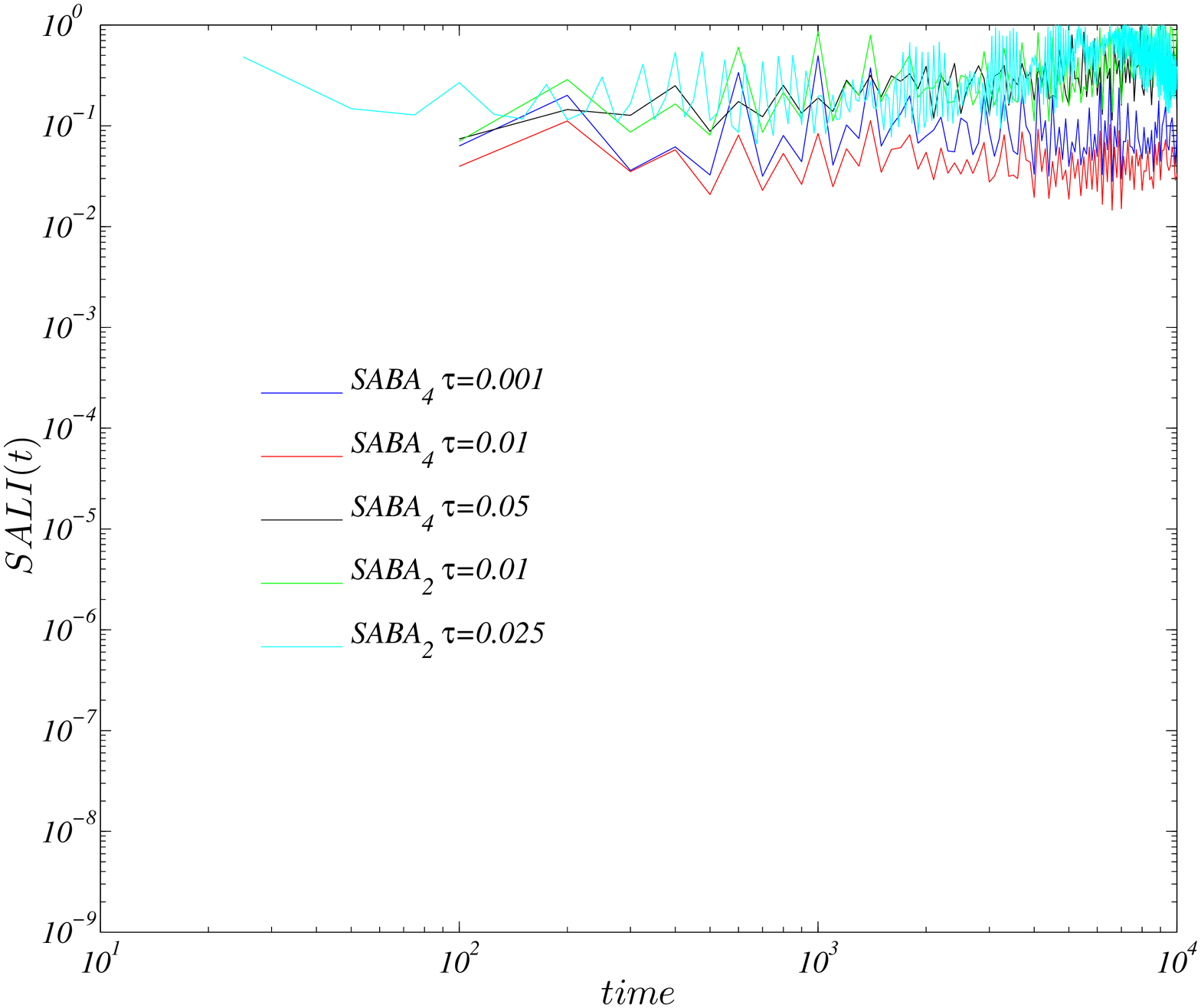}\quad
\includegraphics[width=5cm]{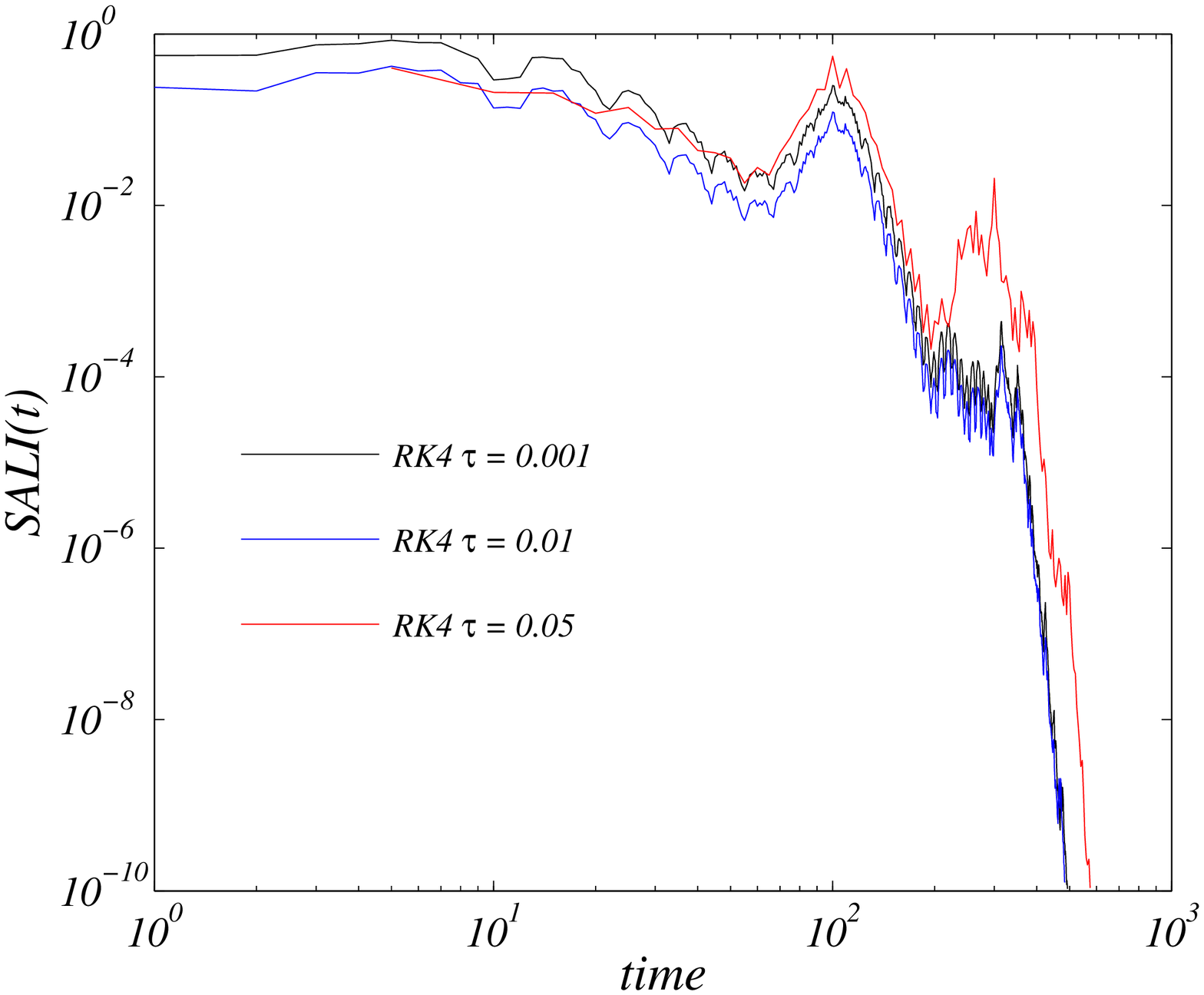}\quad \includegraphics[width=5cm]{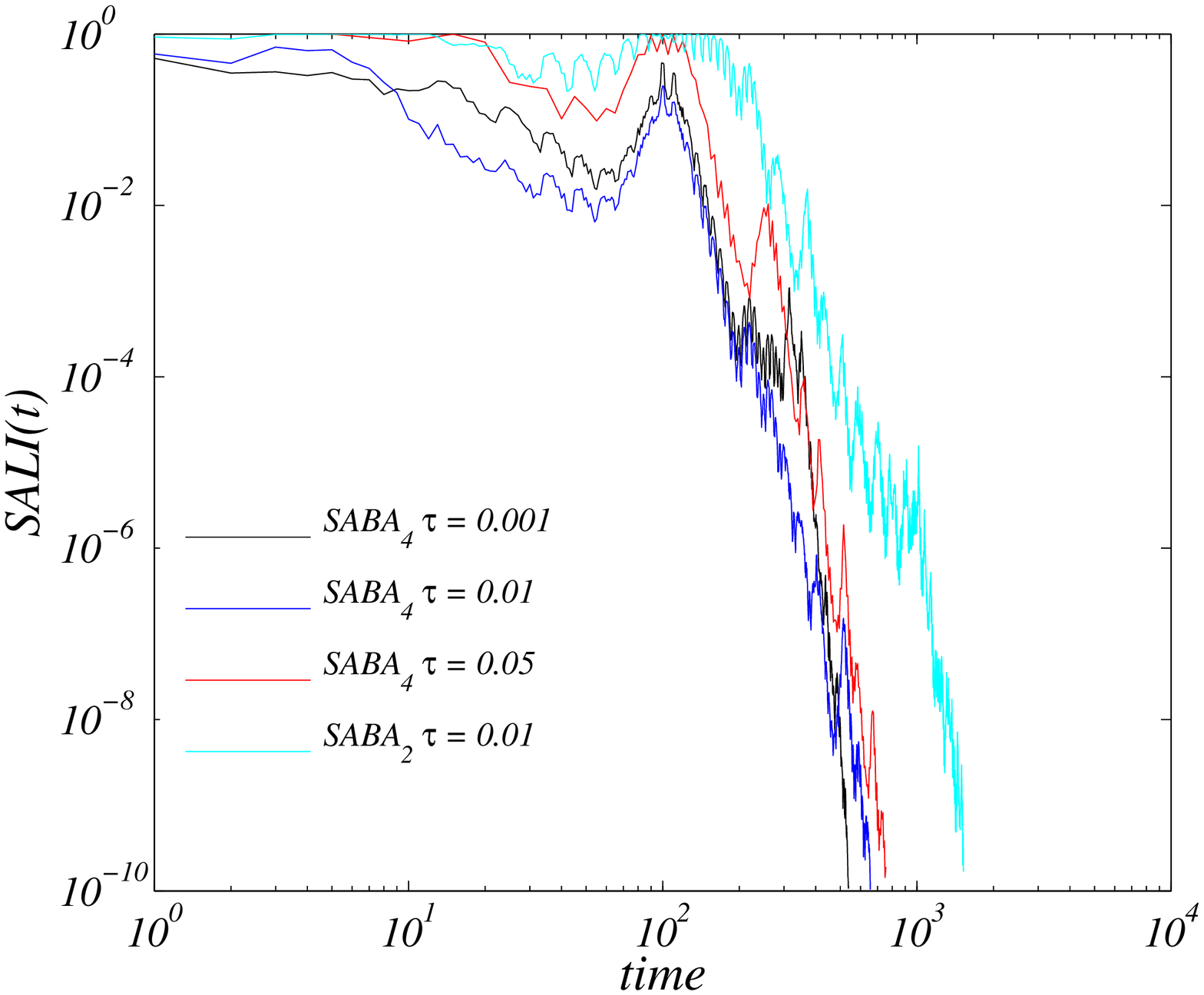}\quad
\end{center} 
\caption{Comparison of integration schemes on the regular orbit $x = 0$, $y = 0.55$, $p_x\sim 0.2417$ (top panels) and $p_y= 0$ and on the chaotic orbit $x = 0$, $y =- 0.016$, $p_x\sim 0.49974$ and $p_y= 0$ (bottom panels). Left panels, \texttt{SALI} computed using non-symplectic $RK4$ for time steps $\tau\in\{ 0.001,0.01,0.05\}$. Right panels, \texttt{SALI} computed using $SABA_4$ for the integration of both the orbit and the deviation vectors, for the same time steps. Both methods determine correctly the character of the orbits independently from the small time step used.}
\label{fig:rk4saba4saba2}
\end{figure}

%\begin{figure}[htbp]
%\begin{center}
%\includegraphics[height=5cm]{22062010_40XV2.eps}\quad \includegraphics[height=5cm]{22062010_30XV2.eps}\quad
%\end{center}
%\caption{Comparison of integration schemes on the chaotic orbit $x = 0$, $y =- 0.016$, $p_x\sim 0.49974$ and $p_y= 0$.  Left panel, \texttt{SALI} computed using non-symplectic $RK4$ while on the right panel, \texttt{SALI} computed using $SABA_4$ and $SABA_2$ for the integration of both the orbit and the deviation vectors, for time steps $\tau\in\{ 0.001,0.1,1.0\}$. Both methods determine correctly the chaotic character of the orbit independently from the time step used.} 
%\label{fig:rk4saba4saba2chao}
%\end{figure}

In Figure~\ref{fig:sali3orbhh}, we report the results of the numerical computation of the \texttt{SALI} for the three generic orbits of
Figure~\ref{fig:hhphasespace}. Since the characteristic period of the orbits
on this energy level, is of the order of $10$ time units, the integration time
span has been fixed to $10^4$ time units. We can observe that the three 
possible dynamical behaviors, regular orbits, chaotic orbits strongly
diffusing and chaotic orbits slowly diffusing, are well identified. Let us
in fact observe that the strong 
chaotic behavior of the red orbit is 
translated to a quick decrease of \texttt{SALI} to zero. On the contrary,
for the black regular orbit, \texttt{SALI} remains bounded away from
zero. The blue orbit has a particular behavior: for quite a long time, up to
$\sim 30\, 000$ time units, this orbit \lq\lq follows closely\rq\rq a
periodic, thus regular, orbit, and \texttt{SALI} remains positive, but
eventually the chaotic character of the orbit manifests and the indicator
correctly goes to zero. Actually the above orbit is close to an unstable
  periodic orbit.

To check the robustness of our method with respect to a non--symplectic one, 
we firstly reanalyze the above three orbits using larger time steps. More
precisely, we numerically computed the \texttt{SALI} indicator using $SABA_2$,
$SABA_4$ and $RK4$ integrators using time steps for which the energy loss is
bounded by sufficient small value, $10^{-3}$. This can be easily computed
using the results of Figure~\ref{fig:energy}, where we report the relative
energy loss, averaged over $50$ randomly chosen orbits, for the integrators
$SABA_4$, $SABA_2$ and $RK4$, as a function of the time steps. For sake of
completeness, we computed the relative energy loss using several integration
time spans. Let us emphasize that, the energy losses for symplectic
integrators are almost the same for all the integration time spans for a fixed
value of $\tau$, on the other hand the non--symplectic scheme $RK4$ exhibits a
strong dependence on the integration time span: the larger the time span is,
the worst is the energy loss for a fixed value of $\tau$. As expected, the
larger the 
time step used, the larger the loss of energy. In the following, we fix the
largest time steps, $\tau$, such that the relative
energy loss is smaller than $10^{-3}$, for an integration
time span equal to $10^4$ time units. Under these assumptions we get:
$\tau_{max}\sim 0.3$ for $SABA_4$, $\tau_{max}\sim 0.16$ for 
$SABA_2$ and $\tau_{max}\sim 0.08$ for $RK4$. As a result for the
H\'enon--Heiles system, our symplectic scheme allows time steps four 
times larger than a non-symplectic one of the same order, as will be shown
hereafter.   

With these maximal realistic time steps in mind, we reanalyzed the three
orbits of Figure~\ref{fig:sali3orbhh} with time steps $\tau\in
\{10^{-3},0.01,0.05\}$ for which the energy loss is small enough, even for the
$RK4$ integrator. All the integration schemes behave almost equally for all
the small used time steps, as it can observed from the results reported in
Figure~\ref{fig:rk4saba4saba2}.

% In the case of the regular orbit, results reported in Figure~\ref{fig:rk4saba4saba2reg} clearly show that the symplectic scheme (right
% panel) outperform the $RK4$ integrator (left panel). Indeed, the orbit
% character can be correctly determined using both a larger step size or a
% smaller integration order. For a chaotic orbit, all the integration schemes
% behave almost equally for all the used time steps, as seen in
% Figure~\ref{fig:rk4saba4saba2chao}. In fact, numerical errors introduced by the
% integrator will not change the orbit character; the only change is the speed
% at which the \texttt{SALI} decreases to zero, hence a different estimation of
% the Lyapunov exponent if one were interested in. As a result, it appears that,
% for increasing time steps, non-symplectic integrators, such as $RK4$, tend to
% show an excessive number of chaotic orbits, while our global
% symplectic integrator for
% the integration of the deviation vector seems able to identify correctly the
% character of the orbits, even for large time steps. 

\begin{figure}[htbp]
\begin{center}
\includegraphics[width=9cm]{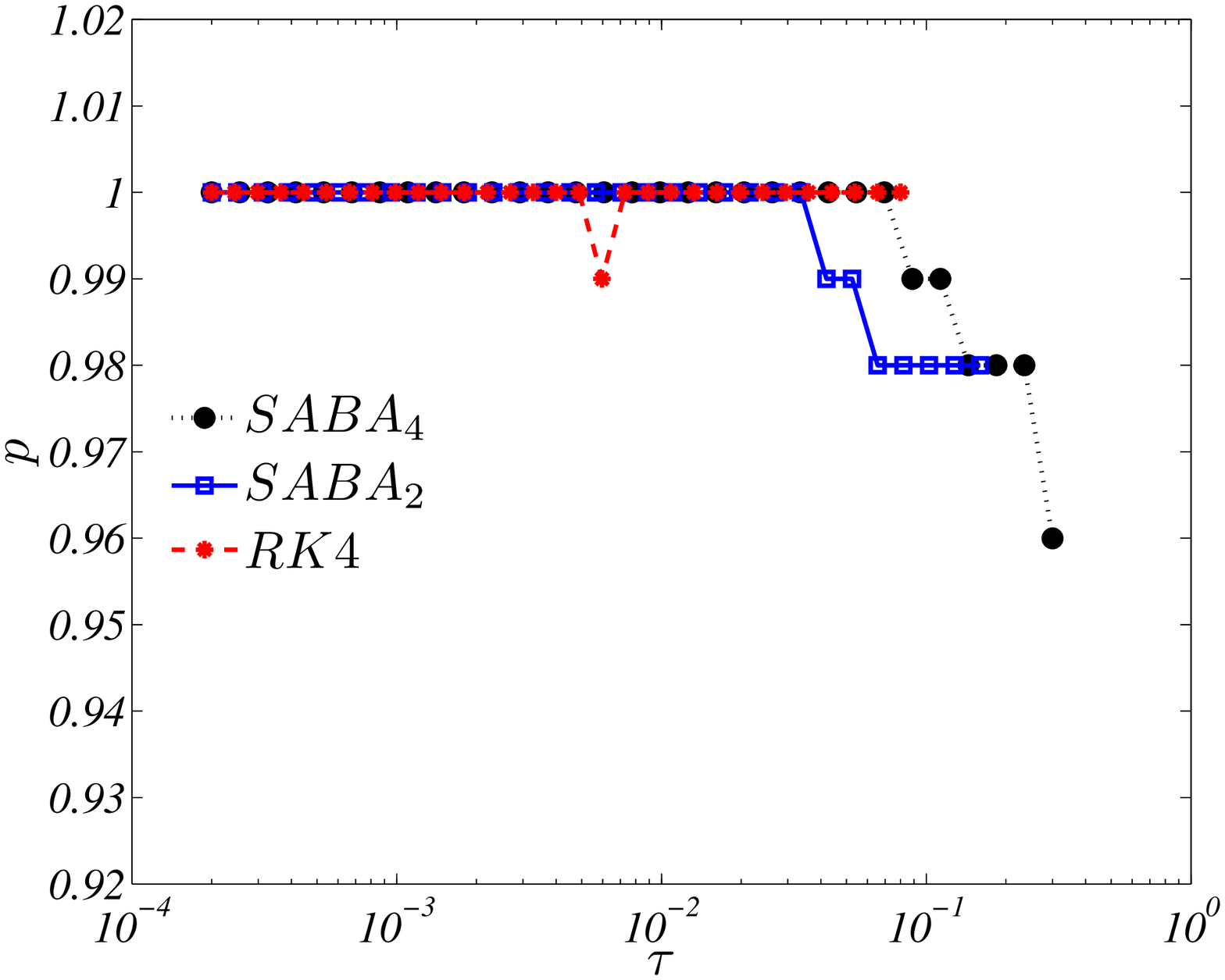}\hfill
\includegraphics[width=9cm]{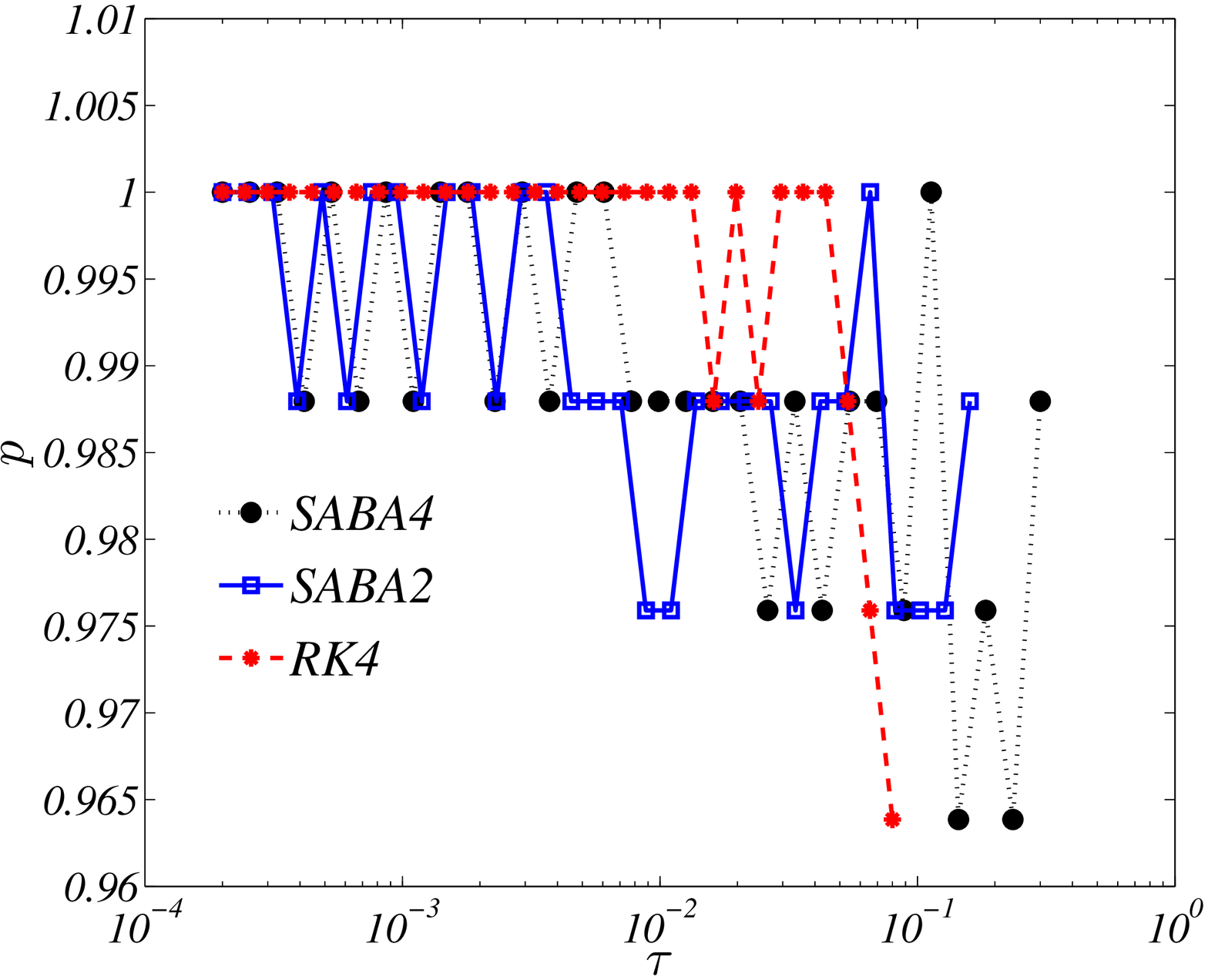}\hfill
\end{center}
\caption{Global comparison between $RK4$ non--symplectic scheme and $SABA_4$,
  $SABA_2$ symplectic schemes for a large portion of the H\'enon--Heiles phase
  space: $N_{reg}=100$ regular orbits and $N_{cha}=100$ chaotic orbits are
  considered. Top panel: Percentage of correctly identified regular orbits for
  increasing time steps ($p=$ correctly identified orbits / total number of
  orbits). Bottom panel: Same as top panel for chaotic orbits.} 
\label{fig:rk4sabanglob}
\end{figure}

The next step of our comparison is to consider a larger portion of phase space
to capture information on the characterization of the global dynamics using
symplectic and non--symplectic integration schemes. We hence
consider $N_{reg}=100$ randomly chosen
regular orbits and $N_{cha}=100$ 
randomly chosen chaotic orbits, whose behavior has been accurately determined using a
sufficiently small step size $\tau$, namely $\tau=10^{-4}$, and a $4^{th}$
order symplectic scheme. Then we compute as a
function of the used step size, how many orbits are correctly 
characterized by the $RK4$ non-symplectic integration,
and by the $SABA_2$ and 
$SABA_4$ symplectic integration for both the orbit and the
deviation vectors.

\begin{figure}[htbp]
\begin{center}
\includegraphics[width=9cm]{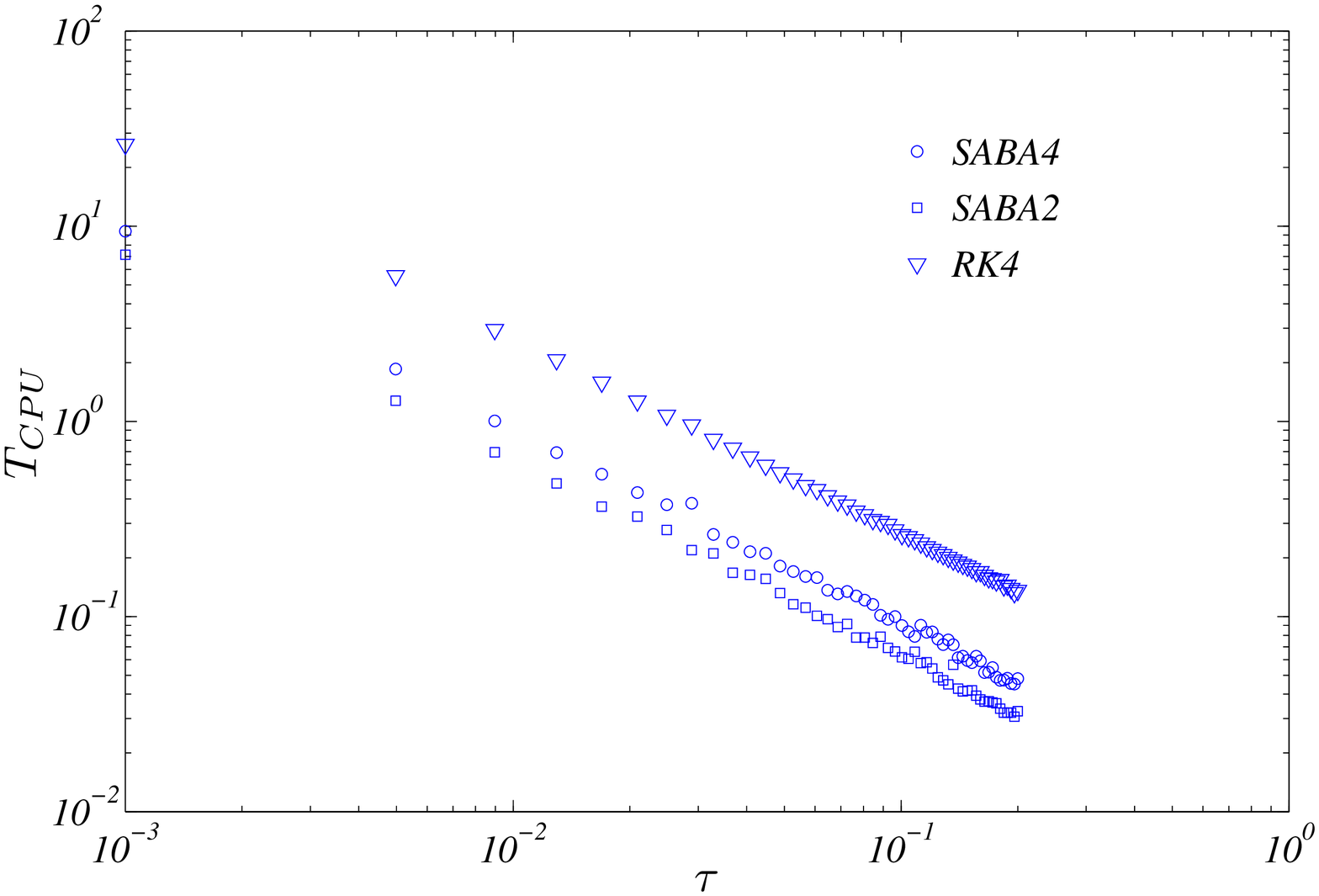}\hfill
\end{center}
\caption{CPU time as a function of the time step $\tau$. Both quantities are
  given in logarithmic scale. Linear best fits (data not shown) provide an
  almost linear decrease of $T_{CPU}$ as a function of $\tau$. The time
  integration span has been fixed to $T_{fin}=10^4$.} 
\label{fig:CPU}
\end{figure}

\begin{figure}[htbp]
\begin{center}
\includegraphics[width=9cm]{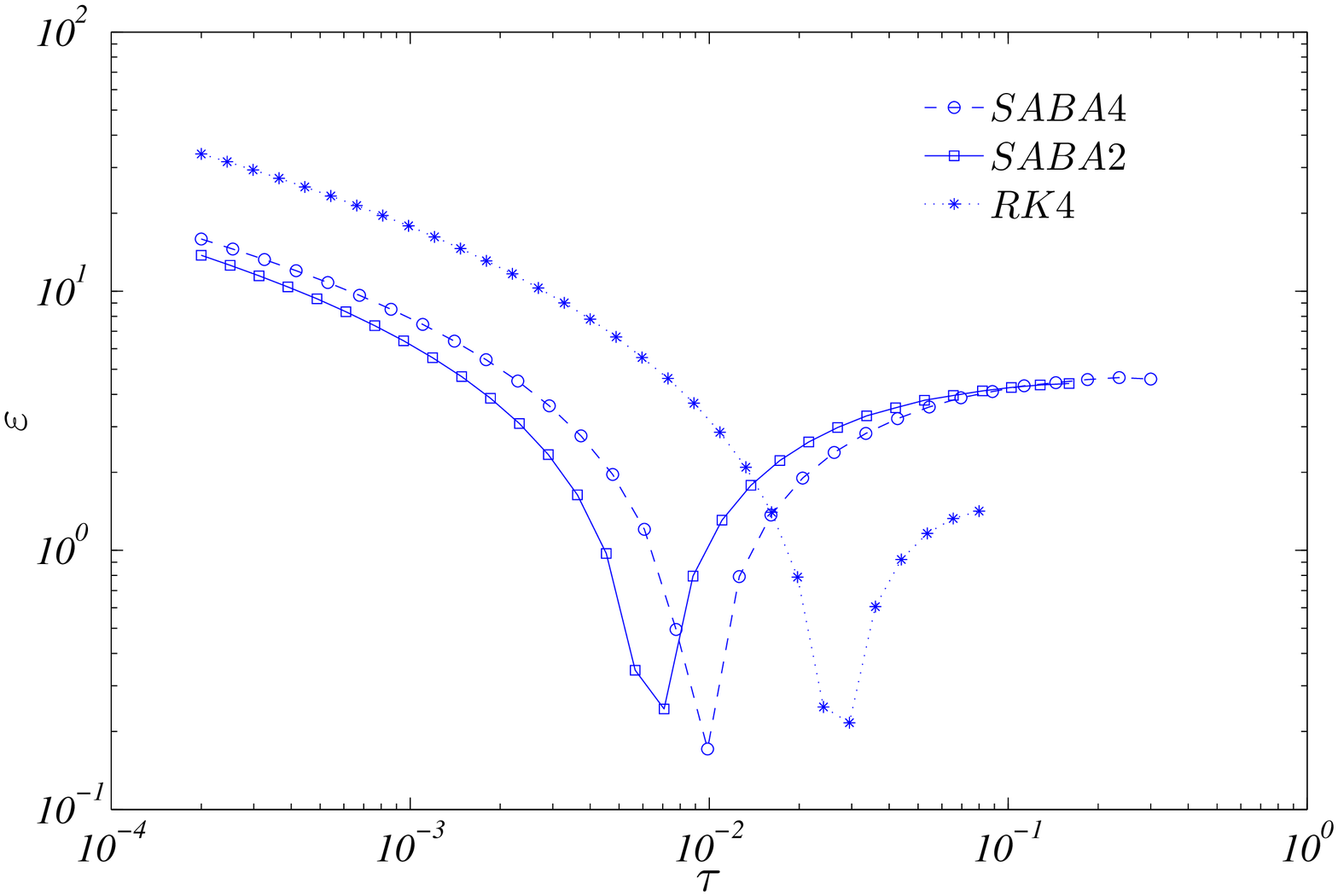}\hfill
\end{center}
\caption{The efficiency $\varepsilon=p |\log_{10}(|\Delta
E/E|)| \,|\log_{10}(T_{CPU})|$ of the global symplectic method ($SABA_4$ and
$SABA_2$) and the non-symplectic one ($RK4$). The scales are logarithmic. See
text for detail.} 
\label{fig:efficiency}
\end{figure}

On the one hand, the results reported in Figure~\ref{fig:rk4sabanglob} clearly
show that, for regular orbits, almost all schemes are able to recover nearly
$100$\% of regular orbits. Concerning the step sizes,
Figure~\ref{fig:rk4sabanglob} shows that $SABA_4$ characterizes correctly the
dynamics of the same percentage of orbits with 4 times larger time steps than
$RK4$, while $SABA_2$ does the same with 2 times larger time steps than
$RK4$, but observe that its order is half the one of $RK4$. On the other hand,
the same conclusion holds for the chaotic 
orbits. Moreover, let us note a general trend of $RK4$ to overestimate the
number of regular orbits when increasing the time step. Thus, our symplectic
scheme is largely better than $RK4$ to characterize regular orbits, regarding
both the used time step and the integration order. Indeed, $SABA_2$ symplectic
integrations reveal to be nearly as reliable as $SABA_4$ ones, especially for
small step sizes. 

In Figure~\ref{fig:CPU} we report the CPU times averaged over fifty randomly
chosen orbits, $T_{CPU}$, as a function of the time steps. One can clearly
observe that the symplectic schemes are faster than $RK4$ using the same step
size, more precisely $SABA_4$ is almost $2.5$ times faster than $RK4$. Once
again this fact illustrates the good numerical performance of our method which
is less time 
consuming in itself but also allows larger time steps, reducing again the
computational time.  

To quantify the {\em efficiency} of our symplectic scheme, we introduce the
following efficiency indicator: $\varepsilon=p |\log_{10}(|\Delta
E/E|)| \,|\log_{10}(T_{CPU})|$, whose dependence as a function of the time step
$\tau$ is represented in Figure~\ref{fig:efficiency}. Let us observe that the
larger is $\varepsilon$, the better is the integration scheme. For small
integration time steps, the non-symplectic $RK4$
scheme dominates, mainly because its energy loss is measly (see
Figure~\ref{fig:energy}). On the other hand, for time steps $\tau$ larger than
$0.1$, the situation is 
reversed: the global symplectic method dominates, as its energy loss and CPU
times are both relatively small (see Figure~\ref{fig:energy} and \ref{fig:CPU}),
that results in a quite large efficiency.

As a result, it appears that the use of a symplectic integrator to
compute the time evolution of the deviation vector can improve the capability of a chaos
indicator such as \texttt{SALI} to determine the character of the orbits in a
Hamiltonian system. Furthermore, we showed that our global symplectic integrator allows larger time steps without energy loss and saves a considerable amount of computation time. This possibility turns out to be essential
for the study of real problems, where the time scales are fixed by
  physical constraints, as for instance in the case of secular resonances in the
N-body celestial mechanics problem, as shown in the next section.

\subsection{Restricted three-body problem}
\label{ssec:3b}

The problem of three celestial bodies interacting with each other
gravitationally is a well-known problem in celestial
mechanics. In the following we will consider the
restricted problem. More precisely one of the bodies, hereby called the Sun, is the largest mass around which the two other bodies are assumed to evolve; the second body, possessing an intermediate mass hereby called Jupiter, orbits around the Sun on a circular orbit, while the third body, hereby called the asteroid, has a negligeable mass and is moving on an inner orbit \lq\lq between \rq\rq the Sun and Jupiter. Assuming this geometry, Kozai \cite{Kozai} showed that a highly inclined 
asteroid perturbated by Jupiter is characterized
by a coupled variation in its eccentricity $e$ and inclination $i$, in such a
way that $H=\sqrt{a(1-e^2)}\cos(i)$ is a constant, $a$ being the asteroid's
semi-major axis. This dynamics is often referred to as {\it Kozai resonance}~(e.g. \cite{Libert}).

\begin{figure}%[htbp]
\begin{center}
\includegraphics[width=13.5cm]{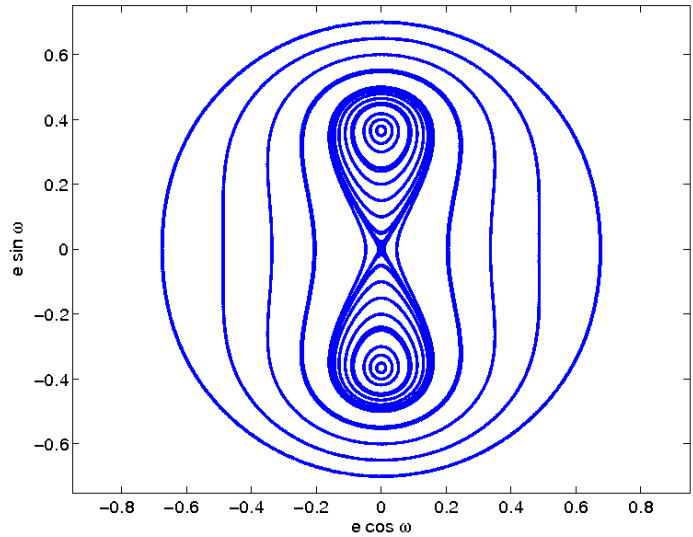}
\end{center}
\caption{Restricted three-body problem phase space, corresponding to $H=0.41833$, and reproducing the Kozai resonance (see the text for further details). Initial conditions for the calculation are the following: $a=0.35$, $a_{giant}=1$, $e_{giant}=0$, $i_{giant}=0$ where the subscript $giant$ refers to the outer giant planet.}
\label{fig:3bphasespace}
\end{figure}

Indeed, the restricted three-body problem can be reduced to two degrees of
freedom after short-period averaging and node reduction (see for
  instance \cite{Malige}). Assuming that the outer giant planet is on a
circular orbit, this problem is integrable, and its dynamics can be
represented on the phase space $(e \cos \omega, e \sin \omega)$ where $\omega$
is the argument of the pericentre of the massless body (see
\cite{Thomas}, \cite{Funk} for more detail). Such a representation
is given in Figure~\ref{fig:3bphasespace} where we plotted
several trajectories of
the small body obtained by the numerical integration of the Hamiltonian
equations associated to the {\it democratic heliocentric} formulation
(see for instance \cite{Duncan}) of the three-body problem: 
\begin{equation}  \label{eq:ham3b}
K(P_i,Q_i)= \sum_{j=1}^2 \left\{\frac{ \|P_j\|^2}{2m_j} - G \frac{m_0m_j}{\|Q_j\|} \right\}+\frac{1}{2m_0} \left\| \sum_{j=1}^2 P_i\right\|^2 - G \sum_{j=1}^2 \sum_{i=1}^{j-1} \frac{m_imj}{\|Q_i-Q_j\|}\, .
\end{equation}
To realize the representation of Figure \ref{fig:3bphasespace}, we used a
$10^{th}$ order SABA integrator with a time step sufficiently small of
$\tau=10^{-3}$ years. This phase space corresponds to $H=0.41833$, which means
an inclination of $45^\circ$ for the circular massless body represented on the
center of the plot. As one can see, for that
large inclination value, circular 
orbit corresponds to an unstable equilibrium point. A separatrix divides the
phase space in three parts: two regions are characterized by the libration of
$\omega$ respectively around $90^\circ$ and
$270^\circ$ and a third one corresponding to a circulation  
of this angle. The two stable equilibria enclosed by the separatrix are
referred to as Kozai equilibria. Hence, a massless body initially on a
circular orbit will suffer from large variations in eccentricity, since its
real motion (short periods included) will stay close to the separatrix of the
reduced problem.  

Let us note that these perturbations on a small body at high inclination are
secular, which means that they operate on extremely long time scales (about
$10,000$ years) compared to the annual orbital periodic variation of the
bodies. Symplectic integrators are recommended for that kind of problem, as
they possess good energy conservation properties and are efficient for large
time steps. Our method for integrating symplectically the deviation vectors is
then particularly suitable for the chaos determination
of such a problem.  

\begin{figure}[htbp]
\begin{center}
\includegraphics[width=10cm]{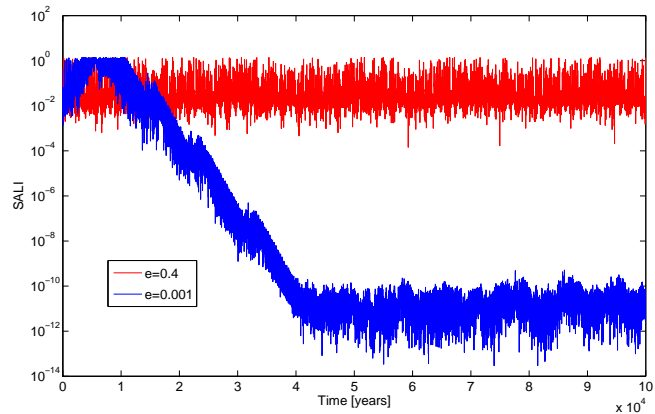}
\end{center}
\caption{\texttt{SALI} characterization of two orbits of Figure \ref{fig:3bphasespace} with a symplectic $SABA_{10}$ scheme: a regular orbit at $e=0.4$ (red) and a chaotic orbit at $e=0.001$ (blue).}
\label{fig:3bSALI}
\end{figure}

In Figure \ref{fig:3bSALI}, we report the results of the numerical computation
of the \texttt{SALI} with $SABA_{10}$ integrator ($\tau=10^{-3}$ years) for a
regular orbit ($e=0.4$, i.e. close to one of the stable Kozai equilibria) and
a chaotic orbit ($e=0.001$, i.e. close to the unstable equilibrium). Our
global symplectic integrator scheme reproduces correctly the expected character of both orbits in less than $40,000$ years. 

\begin{figure}[htbp]
\begin{center}
\includegraphics[width=10cm]{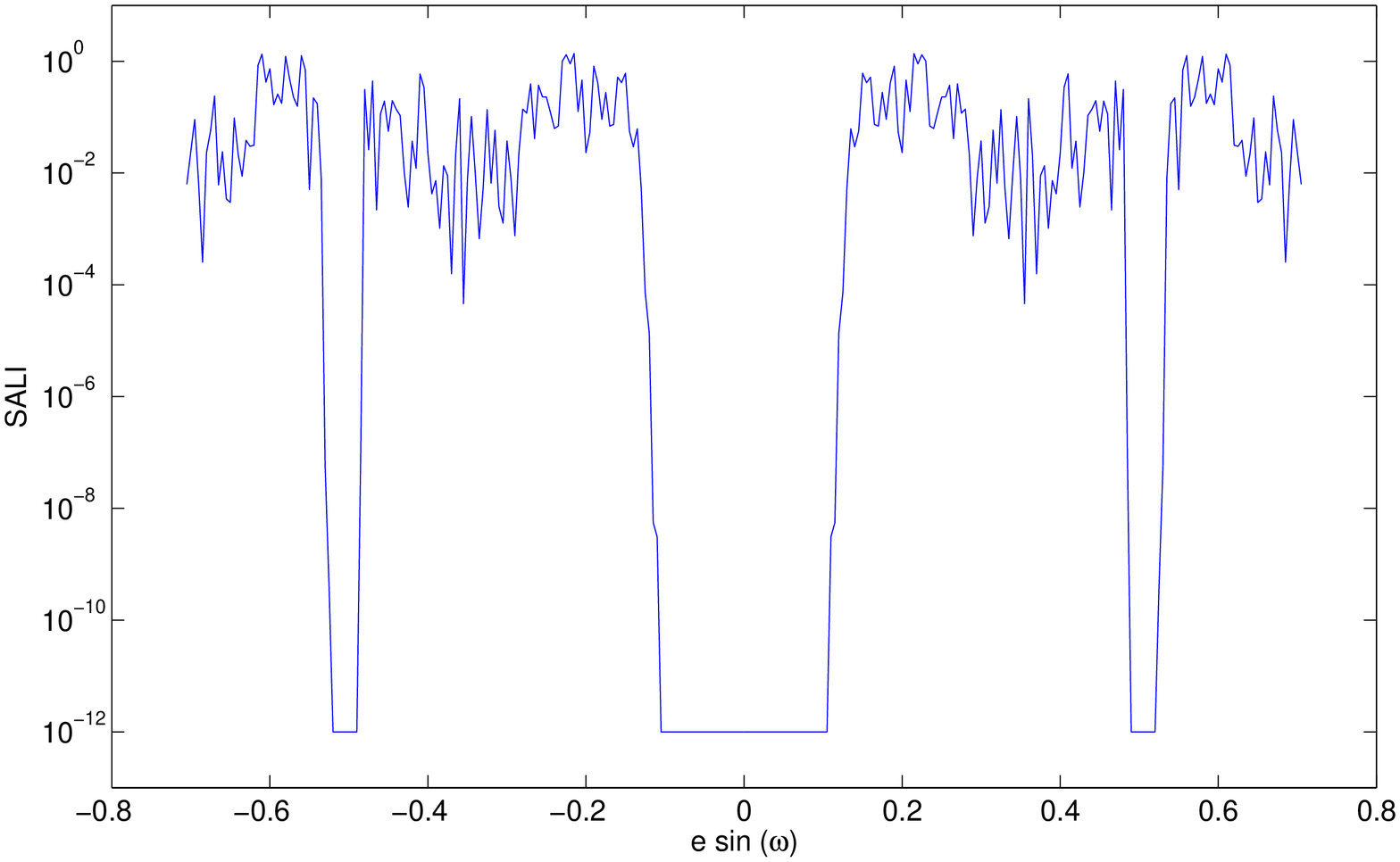}\hfill
\end{center}
\caption{\texttt{SALI} characterization of the global behavior of Figure \ref{fig:3bphasespace} as a function of $e \sin \omega$, after $100,000$ years. In order to avoid useless computation time, \texttt{SALI} values have been fixed to $10^{-12}$ when reaching this threshold. The influence of the separatrix is obvious and well identified by our symplectic $SABA_{10}$ scheme.}
\label{fig:3btranche}
\end{figure}

Concerning the global behavior of the phase space, it can be thoroughly
described by resorting to a chaos indicator technique along a cross section of
the phase space. We choose to represent the \texttt{SALI} values as a function
of the initial values for $e \sin \omega$, as shown in Figure~\ref{fig:3btranche}. The chaos along the separatrix is clearly visible for values of $e
\sin \omega$ close to $0$ and to $0.5$.

\begin{figure}%[htbp]
\begin{center}
\includegraphics[width=6.75cm]{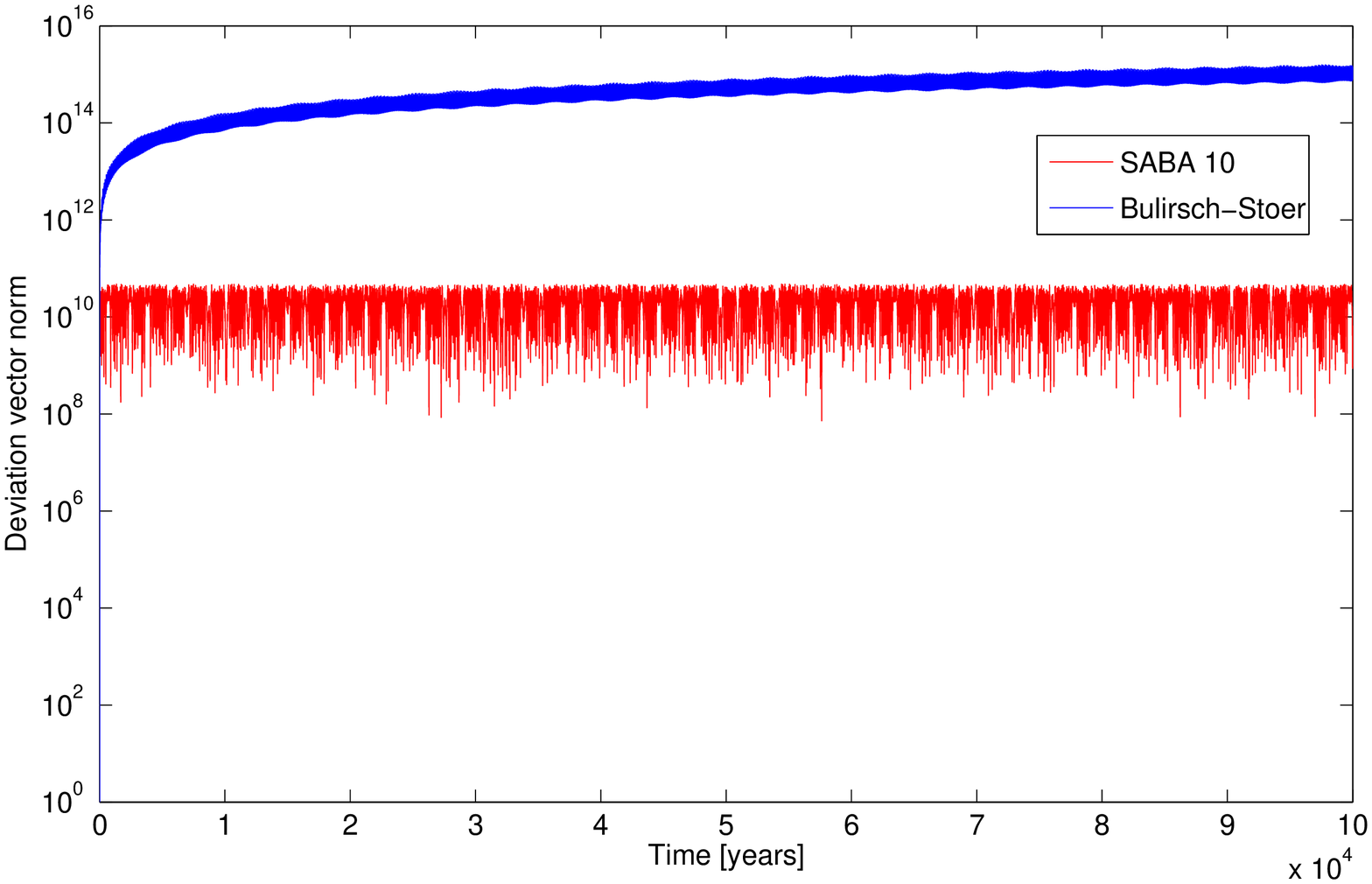}
\includegraphics[width=6.75cm]{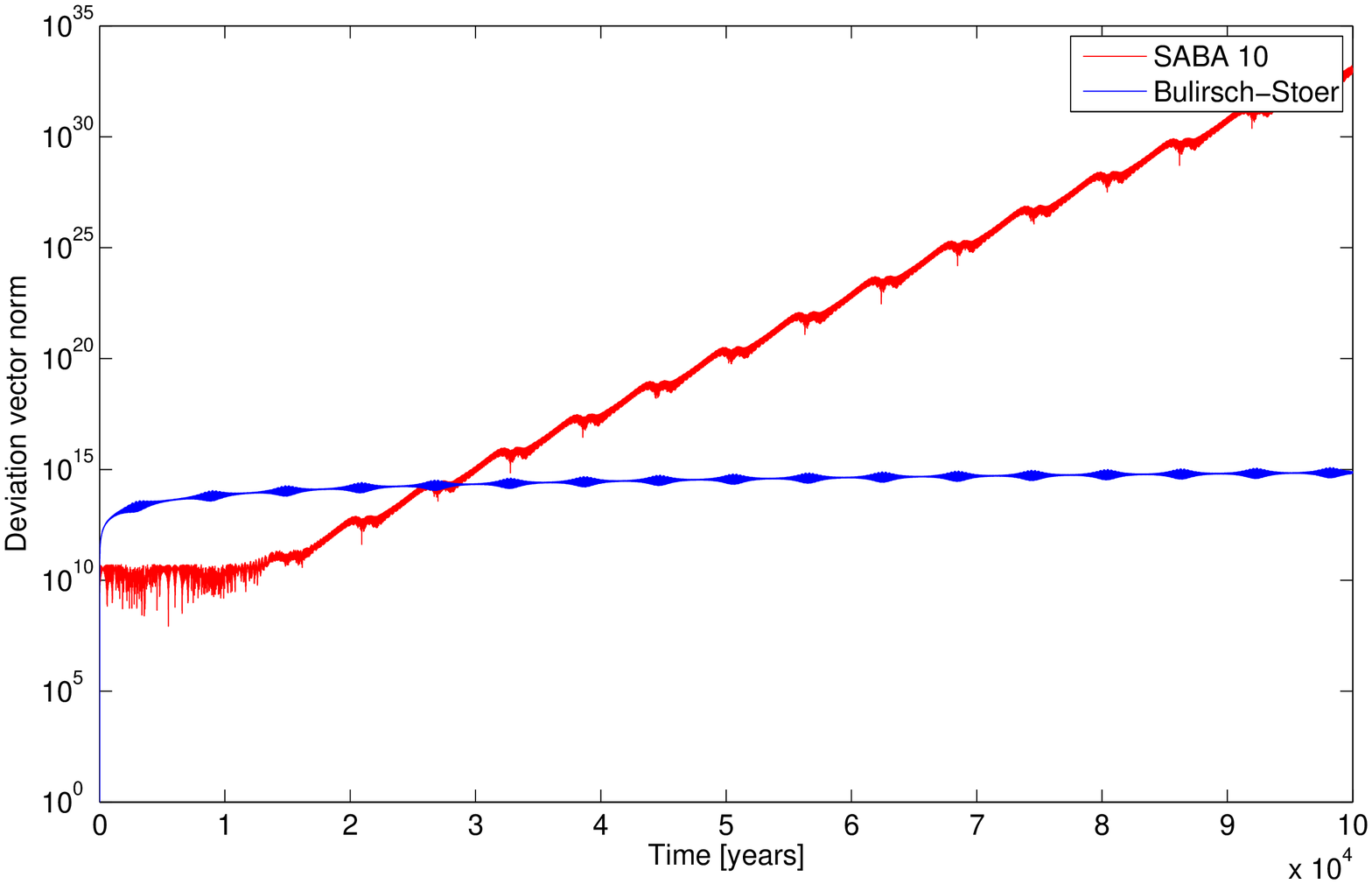}
\end{center}
\caption{Norm of the deviation vector computed with $SABA_{10}$ symplectic scheme (red curves) and Bulirsch-Stoer method (blue curves) for the two orbits of Figure \ref{fig:3bSALI}.}
\label{fig:3bBS}
\end{figure}

In order to compare our method with a non-symplectic scheme, we tried to
reproduce these results with the Bulirsch-Stoer integration method. The first
comment is that this adaptive stepsize method requires computation times up to three times longer than the global symplectic integrator. Secondly,  it appears that this non-symplectic method seems 
unable to characterize correctly the orbits, regardless of the time step
used. Indeed, even if the orbit is correctly described by the Bulirsch-Stoer
method, the deviation vectors do not enable us to distinguish a regular
behavior from a chaotic one. To show this, we reported in Figure \ref{fig:3bBS}
the norm of the same deviation vector integrated with our symplectic scheme
(red curves) and with the Bulirsch-Stoer method (blue curves). Using the 
$SABA_{10}$ method, the norm of the deviation vector grows very rapidly for
the chaotic orbit identified in Figure \ref{fig:3bSALI}. On the contrary, no
significant deviation is observed between the two orbits of
Figure \ref{fig:3bSALI} for the non-symplectic method. Maybe, the integration on
a longer time span with the Bulirsch-Stoer method would eventually reveal the
chaotic behavior of one of these two orbits. Anyway, given that the energy is
not fully conserved, one could not reliably trust this result, even more on long time interval.

\section{Conclusions}
\label{sec:ccl}

In this work, we proposed a new method for the detection of regular and chaotic orbits in Hamiltonian systems, based on the integration of the deviation vectors used in chaos detection techniques, using symplectic algorithms. Our method has been tested on two well-known models, and the results clearly demonstrate that it outperforms non-symplectic ones.

Concerning the H\'enon-Heiles system, it appears that, for large time steps,
non-symplectic integrators tend to detect an excessive number of chaotic
orbits, while the global symplectic integrator is able to identify correctly
the character of nearly all orbits for larger time steps, up to four times
larger than non--symplectic ones. Moreover, due to his symplectic properties,
we showed that our method ensures a very small energy loss even on very long
time spans. Let us emphasize that the possibility to use larger time steps
saves a considerable amount of computation time. 

This possibility turns out to be essential for the study of the Kozai
resonance in the restricted three-body problem, where the secular orbital
changes operate on extremely long time scales. Once again, the influence of
the separatrix of this problem is well identified by our global symplectic integrator. On
the contrary, the Bulirsch-Stoer non-symplectic method, even with smaller time
steps, seems unable to distinguish between regular and chaotic motion, at
least on the same integration time span. A possible reason for this behavior could be the
accumulation of numerical errors introduced by the integrator and a significant energy
loss, disadvantages which are avoided using our symplectic scheme.

We are confident that our findings should be generic for a large class of
Hamiltonian systems. Thus we encourage scientists working on chaos indicators
to perform symplectic integrations of both the orbit and the deviation vectors
using the global symplectic integrator, as proposed in the present paper,
whenever the Hamiltonian is of the form $H(\vec{x})=A(\vec{p})+B(\vec{q})$, or
generically it can be divided into two parts, each one separately
integrable. Computation time and reliability of the results could thus benefit
a lot from this procedure, as we clearly demonstrated above. 

\section*{Acknowledgments}

The work of A.-S. L. is supported by an FNRS Postdoctoral Research
Fellowship. The work of Ch. H. is supported by an FNRS PhD
Fellowship. Numerical simulations were made on the local computing resources
(Cluster URBM-SYSDYN) at the University of Namur (FUNDP, Belgium). The authors
want to thank Ch. Antonopoulos for his constructive suggestions that we used
to improve a previous version of our work.

\section*{References}

%\bibliographystyle{elsarticle-harv}

\end{document}